\documentclass[aps,prl,preprint]{revtex4-1}
\usepackage{amsmath}
\usepackage{graphicx}
\usepackage{textcomp}
\usepackage[hidelinks]{hyperref}

\begin{document}

\title{Time-of-Flight Electron Energy Loss Spectroscopy by Longitudinal Phase Space Manipulation with Microwave Cavities}
\author{W. Verhoeven}
\author{J. F. M. van Rens}
\author{W. F. Toonen}
\affiliation{Department of Applied Physics, Coherence and Quantum Technology Group, Eindhoven University of Technology, P.O. Box 513, 5600 MB Eindhoven, the Netherlands}
\author{E. R. Kieft}
\affiliation{Thermo Fisher Scientific, Achtseweg Noord 5, 5651 GG Eindhoven, the Netherlands}
\author{P. H. A. Mutsaers}
\author{O. J. Luiten}
\email{o.j.luiten@tue.nl}
\affiliation{Department of Applied Physics, Coherence and Quantum Technology Group, Eindhoven University of Technology, P.O. Box 513, 5600 MB Eindhoven, the Netherlands}
\date{\today}

\begin{abstract}
The possibility to perform high-resolution time-resolved electron energy loss spectroscopy has the potential to impact a broad range of research fields. Resolving small energy losses with ultrashort electron pulses, however, is an enormous challenge due to the low average brightness of a pulsed beam. In this letter, we propose to use time-of-flight measurements combined with longitudinal phase space manipulation using resonant microwave cavities. This allows for both an accurate detection of energy losses with a high current throughput, and efficient monochromation. First, a proof-of-principle experiment is presented, showing that with the incorporation of a compression cavity the flight time resolution can be improved significantly. Then, it is shown through simulations that by adding a cavity-based monochromation technique, a full-width-at-half-maximum energy resolution of 22~meV can be achieved with 3.1~ps pulses at a beam energy of 30~keV with currently available technology. By combining state-of-the-art energy resolutions with a pulsed electron beam, the technique proposed here opens up the way to detecting short-lived excitations within the regime of highly collective physics.
\end{abstract}

\maketitle

\emph{Introduction.}---Over the last decade, electron energy loss spectroscopy (EELS) inside a transmission electron microscope has progressed towards the few meV regime, thus enabling the detection of fundamental excitations with an atom sized probe. With the availability of monochromators on commercial microscopes, EELS has become a routine technique for nanophotonic research~\cite{Schaffer2009,Rossouw2011,Nicoletti2011,Rossouw2013}. Using state-of-the art monochromators, aberration correctors, and EELS detectors, sub-10 meV resolutions have been achieved, which can be used for phonon spectroscopy~\cite{Krivanek2014,Lagos2017,Idrobo2018}. Owing to these advances, transmission EELS can play a pivotal role in understanding some of the greatest problems in modern material science such as strange metal phases and high-$T_c$ superconductivity~\cite{Cvetkovic2008}.

In a very exciting recent development, EELS is combined with high brightness ultrashort electron pulses~\cite{Carbone2008,Flannigan2012,Feist2017}, making it an even more powerful and diverse technique. Using time-resolved studies of the core-loss spectrum, structural changes in a sample can be measured with sub-ps resolutions~\cite{VanDerVeen2015}. Furthermore, interactions between femtosecond lasers and a sample can be measured through the so-called PINEM effect~\cite{Barwick2009,Feist2015}, which has been used to image photonic structures~\cite{Piazza2015} and the dynamics of surface plasmon polaritons~\cite{Lummen2016}. However, the energy resolution of typical time-resolved EELS setups is limited to 1--2~eV, due to the fact that conventional EELS methods require significant spatial filtering of the electron beam.

Because of the limitations of conventional techniques, time-of-flight (ToF) methods are emerging as an alternative, in which electron energies are derived from the flight time across a drift space. These methods put much less demands on the transverse beam quality, allowing for the detection of energy losses without requiring spatial filtering~\cite{Gliserin2016,Verhoeven2016}. In this paper, we will demonstrate that ToF methods can be improved significantly to the level of state-of-the-art conventional methods by employing time-dependent electromagnetic fields to manipulate the longitudinal phase space distribution of an electron beam. These fields are generated using microwave cavities oscillating either in the TM$_{010}$ mode, which has a longitudinal electric field used for time-dependent (de)acceleration, or in the TM$_{110}$ mode, which has a transverse magnetic field used to periodically deflect a beam. In the method proposed here four cavities are used to first create pulses, then to modify the energy spread and temporal length at the sample as desired, and finally to detect energy losses with high accuracy and without any loss of current. Through simulations it will be shown that 22~meV energy resolution can be achieved combined with 3.1~ps temporal resolution. Alternatively, if a better temporal resolution is required, pulses can be compressed at the expense of energy resolution. We show that in the same setup a 2.8~eV energy resolution can be used in combination with 20~fs pulses, or any combination in between with a similar longitudinal emittance, providing a large degree of flexibility.

\begin{figure*}[tbp]
\includegraphics[width=\textwidth]{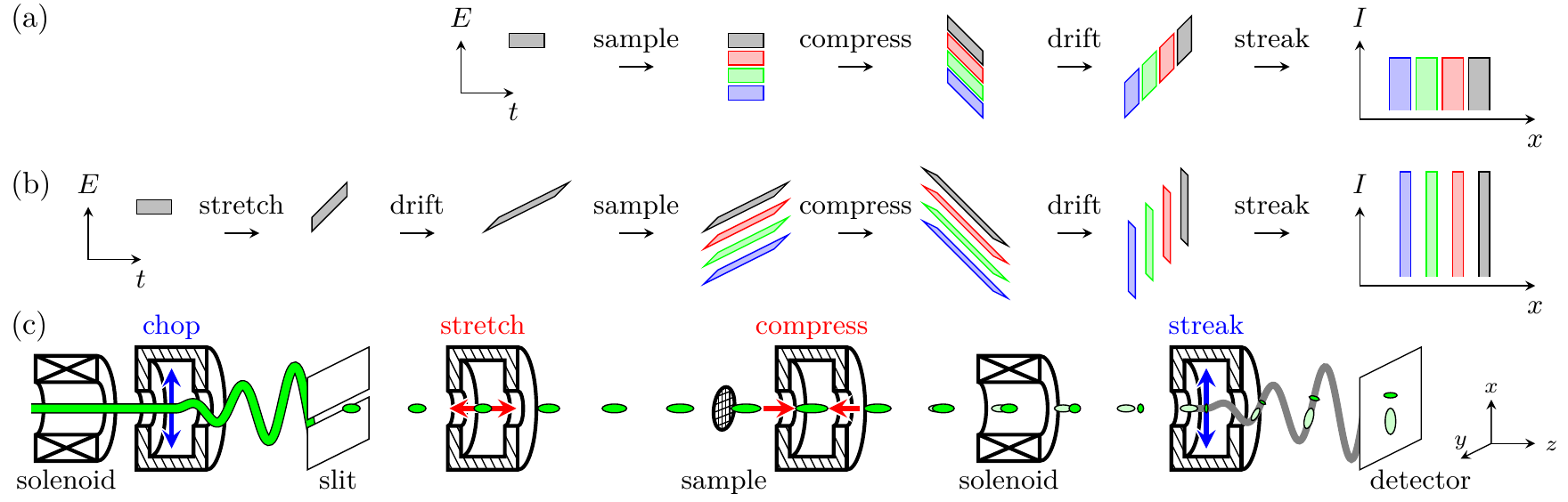}
\caption{(a)~Proposed modified time-of-flight detection scheme, in which different energies are dispersed by means of compression. Shown here are phase space diagrams at each step, followed by the measured intensity after streaking the final temporal distribution onto a detector. (b)~Similar, but now with pulse stretching added in front of the sample. (c)~Proposed four cavity setup, in which pulses can be created, modulated, and detected with a high accuracy.}\label{fig:setup}
\end{figure*}

\emph{Method.}---The idea behind a ToF measurement is to disperse different electron energies temporally throughout a drift space. However, to achieve meV energy resolutions, either extremely short pulse lengths or impractically long drift times are required. Therefore, we propose a modified ToF setup which removes this limitation by incorporating a TM$_{010}$ compression cavity. The idea behind the modified ToF measurement is shown in Fig.~\ref{fig:setup}(a). Shown here are longitudinal phase space diagrams of an electron pulse that interacts with a sample, after which it is compressed. Energy losses induced by the sample are shown in colors, zero-loss electrons are shown in gray. Using a time-dependent electric field that can be generated with a compression cavity, a linear correlation between longitudinal position and velocity can be added to the electrons, causing the pulses to compress throughout the drift space. At the longitudinal focus position of the compression cavity, the different energy losses induced by the sample become temporally separated, independent of the initial pulse length or drift space, leading to a significantly improved flight time resolution. The energy difference $\Delta E$ is then related to a difference in flight time $\Delta t$ through
\begin{equation}
\frac{\Delta E}{E_0}=(\gamma_0+\gamma_0^2)\frac{v_z\Delta t}{L}\,,\label{eq:loss}
\end{equation}
with $E_0$ the initial kinetic energy, $v_z$ the average velocity, $\gamma_0$ the average Lorentz factor, and $L$ the distance from the compression cavity to the longitudinal focus.

Assuming perfect compression and detection, the energy resolution depends solely on the initial energy spread of the electrons. To decrease this energy spread, we propose to use a second TM$_{010}$ compression cavity placed before the sample to stretch the pulses. Figure~\ref{fig:setup}(b) shows again the longitudinal phase space diagrams at each step, but now including the pulse stretching. The longitudinal phase space area, or, equivalently, the longitudinal emittance, is conserved after drift or (de)focusing. Therefore, the pulses can be monochromated by stretching them. Although the total energy spread at the sample is larger in this way, it is correlated with arrival time. The uncorrelated energy spread, however, has decreased, so that energy losses smaller than the initial energy spread of the source become separated in phase space. As an optical analogy, the compressing cavity works as a lens, making a longitudinal image of the different velocities in its backfocal plane. By first defocusing the pulse with the stretching cavity, the numerical aperture increases, and the longitudinal image becomes sharper.

By combining this with a TM$_{110}$ streak cavity to create ultrashort pulses with a high beam quality~\cite{VanRens2017,Verhoeven2018} and a second streak cavity to detect the flight times~\cite{Oudheusden2010,Maxson2017}, pulse creation, modulation, and detection can easily be synchronized. A temporal resolution of a few fs has already been demonstrated experimentally with microwave cavities~\cite{Maxson2017}, which would translate into a few tens of meV energy resolution for a 30~keV beam and a drift space of 1~m. Figure~\ref{fig:setup}(c) shows the total setup with four cavities.

Stretching the pulses comes at the cost of temporal resolution. Alternatively, pulses can be compressed towards the sample for a better temporal resolution at the cost of energy resolution. This provides a large flexibility, allowing for the longitudinal phase space distribution to be optimized for each specific measurement. As with any method, the combination of resolutions achievable depends on the initial longitudinal emittance of the pulses.

\emph{Experimental results.}---In order to demonstrate the modified ToF detection scheme, a proof-of-principle measurement was performed by adapting the ToF setup previously reported in Ref.~\cite{Verhoeven2016}. In this setup, a continuous beam from a Philips XL-30 SEM with a tungsten filament gun is chopped into pulses using a dielectrically loaded TM$_{110}$ cavity~\cite{Lassise2012}. The electrons then interact with a standard calibration sample consisting of a polycrystalline gold layer deposited on a carbon film.

\begin{figure}[tbp]
\includegraphics{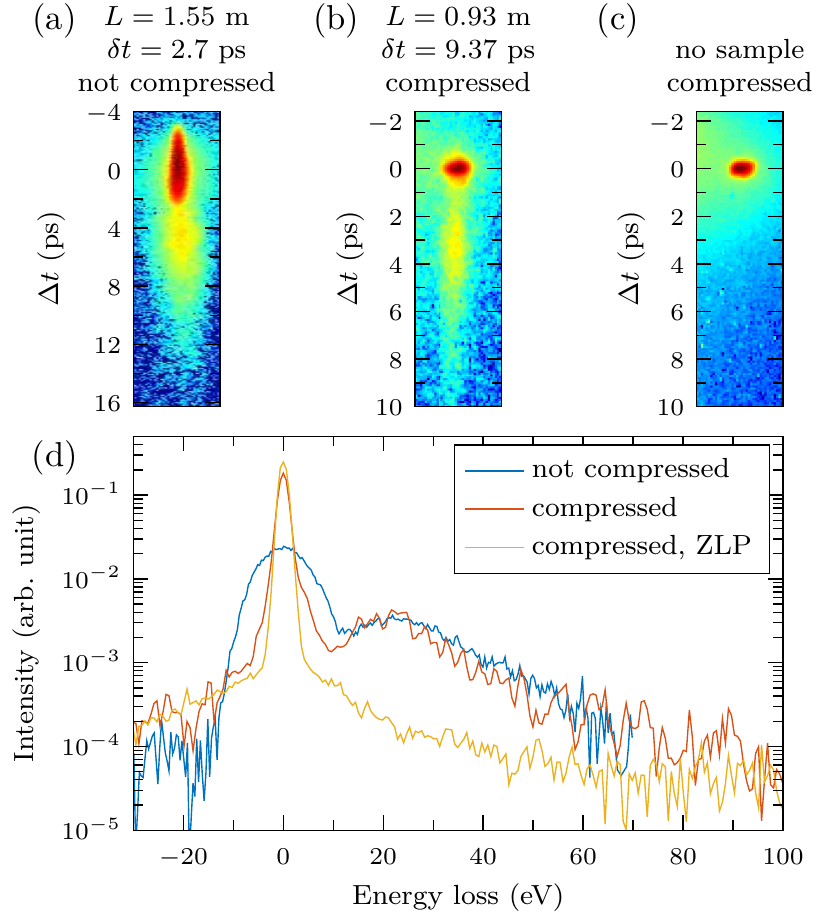}
\caption{(a) ToF measurement without compression as was reported previously in Ref.~\cite{Verhoeven2016}, compared to (b)~a ToF measurement with compression, color coded on a logarithmic scale. (c)~Compressed pulses without a sample inserted. (d)~EELS spectra for all three measurements.}\label{fig:measurements}
\end{figure}

In the previously reported experiment, ToF measurements were performed with 2.7~ps electron pulses over a distance of 1.55~m. The resulting detector signal is shown in Fig.~\ref{fig:measurements}(a). For the proof-of-principle experiment, a compression cavity was added behind the sample, and longer pulses were used. These pulses are compressed from $9.37\pm0.02$~ps to $354\pm7$~fs over a distance of $L=0.93\pm0.01$~m.

Figures \ref{fig:measurements}(b) and (c) show the detector signal of the compressed pulse on a logarithmic color scale, with and without a sample inserted in front of the compression cavity respectively. This shows that indeed flight times can be resolved with a resolution determined by the compressed pulse length, and not the initial pulse length, which would otherwise completely obscure the differences in arrival time associated with the plasmon losses, demonstrating the drastic improvement in the flight time resolution that can be achieved by adding a compression cavity.

Figure \ref{fig:measurements}(d) shows the three EEL spectra calculated using Eq.~\eqref{eq:loss}. All curves are normalized by area. Good agreement is found in the plasmon loss region. With compression, an energy resolution is found of $\delta E_\textsc{fwhm}=2.38\pm0.07$~eV, significantly better than the previously reported resolution of $12\pm2$~eV. The resolution achieved with the compressed pulses is in line with the expectations for a tungsten filament gun, which typically has energy spread of 1.5--3~eV~\cite{ReimerTEM}.

\emph{Simulations.}---Although the current setup is not ideal for time-resolved spectroscopy due to the relatively low brightness and high energy spread of the electron gun, it shows that the modified ToF scheme works as expected. In order to investigate the performance of the technique when monochromation is added, charged particle tracking simulations were performed on the setup proposed in Fig.~\ref{fig:setup}(c) using the General Particle Tracer code (GPT)~\cite{gpt}, which tracks the particle positions using a fifth order embedded Runge-Kutta solver. Cavities are modeled using realistic on-axis field profiles. Off-axis fields are calculated using a fifth order field expansion, yielding an accurate representation of the fields in a cavity, including fringe fields. Solenoids are modeled by an analytical solution that fits well with the field of the solenoid used in the experiments, resulting in realistic spherical and chromatic aberrations.

\begin{figure}
\includegraphics{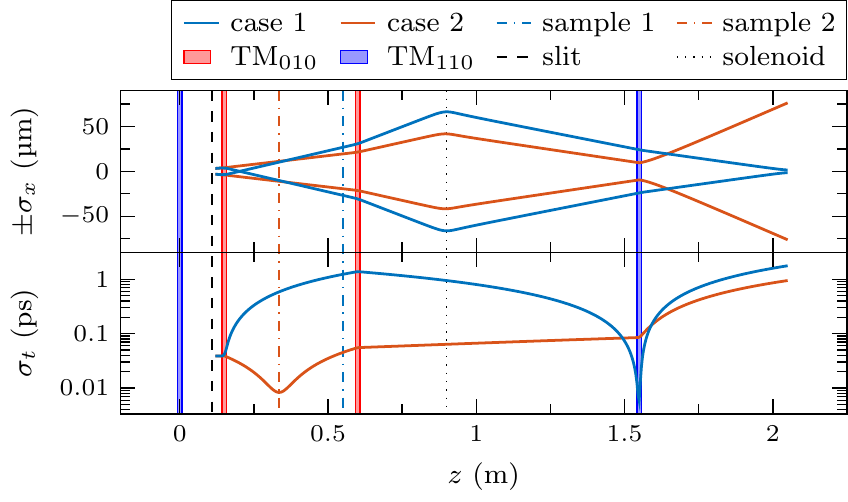}
\caption{Transverse ($\sigma_x$) and longitudinal ($\sigma_t$) beam size throughout the setup when either a good energy resolution is required (case 1), or a good temporal resolution (case 2). All elements in the simulation are also indicated. Note that the optimal sample position is different for case 1 and case 2.}\label{fig:beam}
\end{figure}

The simulations start at the exit of a 30~keV SEM column with a continuous beam, which is assumed to have a transverse emittance of $\varepsilon_{n,x,y}=3$~pm~rad, with an energy spread of $\delta E_\textsc{fwhm}=0.5$~eV~\cite{ReimerTEM}. Shown in Fig.~\ref{fig:beam} are all the elements used in the simulations, and the transverse and longitudinal RMS size of the beam after passing the slit. Shown here are two different situations, in which either an optimal energy resolution is required (case 1), or an optimal temporal resolution (case 2). To facilitate the compression in the second case, the sample is placed at a different position. For all other simulations, the sample is placed near the second compression cavity.

By focusing the continuous beam at the center of the first cavity the emittance of the beam is conserved~\cite{VanRens2017,Verhoeven2018}. In the simulations, this is done with a half-angle of 50~\textmu{}rad, after which the beam is deflected at a magnetic field amplitude of 3~mT over a 10~\textmu{}m slit, placed at a distance of 10~cm. This yields pulses with a RMS pulse length of $\sigma_t=38.8$~fs, and an RMS energy spread of $\sigma_E=0.29$~eV, or $\delta E_\textsc{fwhm}=0.68$~eV.

The chopped pulses are then sent through the rest of the setup, where they are either stretched or compressed when arriving at the sample. After interaction with the sample, the pulses are compressed such that the pulse length is minimal at the center of the final TM$_{110}$ streak cavity. Due to fringe fields of the TM$_{010}$ cavities, the beam is transversely defocused in the compression phase and focused in the stretching phase~\cite{Pasmans2013}, but this can be counteracted with the solenoid, used to focus the beam on the detector. Although no requirements on the beam size at the sample have currently been set, additional magnetic lenses can be used to accomplish this.

\begin{figure}
\includegraphics{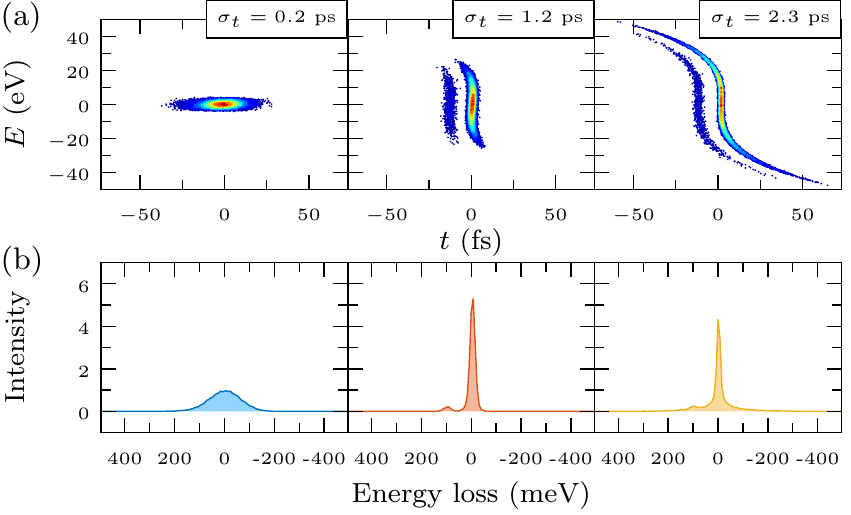}
\caption{(a) Longitudinal phase space of the compressed pulses color coded by density, for three different pulse lengths at the sample. An energy loss of 100 meV is added at the sample for 5\% of the particles. (b)~Resulting EEL spectrum after streaking each distribution on the detector}\label{fig:details}
\end{figure}

Figure~\ref{fig:details}(a) shows the compressed phase space at $z=1.55$~m for three different pulse lengths at the sample. As a demonstration, a fictitious loss peak is added in the sample at 100~meV, which scatters 5\% of the particles. As the flight time from the compression cavity to the longitudinal focus remains constant, the difference in arrival time between the unscattered and the scattered electrons is fixed for all three situations. By first stretching the pulses, the compressed pulse length decreases, resulting in a better separation in phase space of the scattered electrons. For the longest pulse length, however, the temporal separation deteriorates due aberrations of the compression cavity.

Figure~\ref{fig:details}(b) shows the resulting spectra after streaking the distributions on the detector with the final streak cavity. For the streaking cavity a magnetic field amplitude of 6~mT is used, resulting in an energy dispersion of 0.127 \textmu{}m\,meV$^{-1}$ at the detector, which is more than enough to ensure that the detected image is not distorted by \emph{e.g.}\ chromatic aberrations of the solenoid or the transverse beam quality. As expected, the resolution improves by stretching the pulses up to the point where longitudinal aberrations take over.

\begin{figure}
\includegraphics{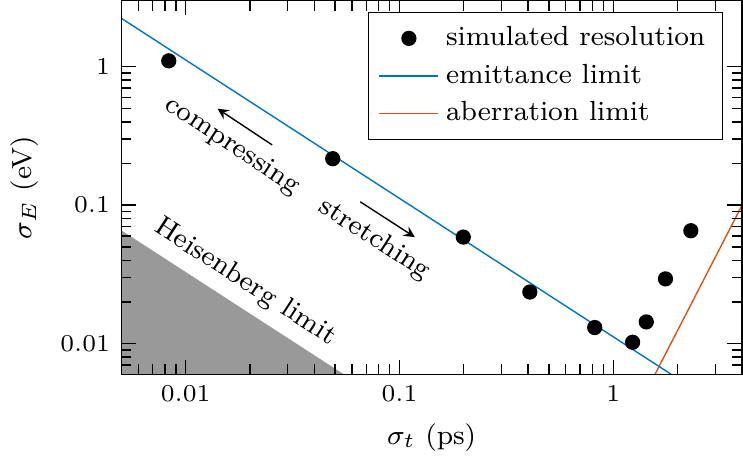}
\caption{Simulated RMS energy resolution as a function of the RMS temporal spread at the sample. Also shown are the theoretical limits, based on the initial longitudinal emittance and the aberrations of the compressing cavity as given by Eq.~\eqref{eq:aberrations}.}\label{fig:resolutions}
\end{figure}

Shown in Fig.~\ref{fig:resolutions} is the combined RMS temporal spread at the sample and RMS energy resolution at the detector, indicated by the black circles, for varying fields in the stretching cavity. As expected, a combination of resolutions can be reached predominantly limited by the initial longitudinal emittance, up to the point where aberrations from the TM$_\text{010}$ cavities take over, after which the longitudinal focus scales proportional to $\sigma_t^3$. Both these limits are indicated by the solid lines. The longitudinal aberrations arise from nonlinearities in the compression field, which is sinusoidal with arrival time to good approximation~\cite{Pasmans2013}. Assuming the compression of a pulse with a Gaussian initial temporal distribution with a spread $\sigma_t$, the RMS spread added to the energy spectrum due to the aberrations of the second compression cavity is given by
\begin{equation}
\sigma_{E,\text{ab}}=\sqrt{\frac{5}{12}}\omega^2(\gamma_0+\gamma_0^2)\frac{E_0}{t_0}\sigma_{t}^3\,,\label{eq:aberrations}
\end{equation}
with $\omega$ the angular frequency of the field in the cavity. The aberrations arise sooner in the simulations due to the additional nonlinearities induced by the stretching cavity. Optimizing the placement of all elements might shift the resolutions closer to the limit of Eq.~\eqref{eq:aberrations}, although a significant improvement is not expected.

The lowest energy resolution found in the simulations is $\delta E_\textsc{fwhm}=22$~meV for pulses with a pulse length of $\delta t_\textsc{fwhm}=3.1$~ps at the sample. Alteratively, by compressing the pulses a temporal resolution of $\delta t_\textsc{fwhm}=20$~fs is found, with an energy resolution of $\delta E_\textsc{fwhm}=2.8$~eV. The longitudinal emittance of these pulses is only a factor 34 above the Heisenberg energy-time uncertainty limit. Interestingly, the longitudinal emittance can be decreased further, at the cost of current. This can be achieved by standard monochromation techniques, or by chopping the stretched pulse a second time. If the setup can be made stable enough, this is a viable prospective.

\emph{Discussion and outlook.}---We conclude that ToF methods combined with longitudinal phase space manipulation employing microwave cavities enables state-of-the-art energy resolutions with ultrashort pulsed electron beams. This can be achieved with relatively simple setups, without the need for spatial filtering. As pulsed electron beam capabilities are pushed further towards the attosecond regime~\cite{Baum2018}, ToF methods will become ever more powerful.

The method proposed here can easily be combined with diffraction techniques, allowing for complete dispersion curves to be measured with a high accuracy as a function of time. Furthermore, the method can be improved by adding longitudinal aberration correction, for which different methods are already available~\cite{Floettmann2014,Zeitler2015}. Pulses can then be stretched further, bringing few meV resolution within reach. This will open up the way to detecting fundamental excitations on all relevant length, time and energy scales.

\begin{acknowledgments}
This work is part of an Industrial Partnership Programme of the Foundation for Fundamental Research on Matter (FOM), which is part of the Netherlands Organisation for Scientific Research (NWO), in collaboration with Thermo Fisher Scientific. The authors would like to thank E.~H.~Rietman, I.~Koole, H.~A.~van Doorn, and A.~H.~Kemper for their invaluable technical support.
\end{acknowledgments}

\bibliography{bibFileV2}
\end{document}